# Spin-orbit interaction of photons and fine splitting of levels in ring dielectric resonator

K.Yu. Bliokh[1,2*] and D.Yu. Frolov[3]

[1]*Institute of Radio Astronomy, 4 Krasnoznamyonnaya st., Kharkov, 61002, Ukraine*
[2]*Department of Physics, Bar-Ilan University, Ramat Gan, 52900, Israel*
[3]*Kharkov National University, 4 Svobody sq., Kharkov, 61077, Ukraine*

We consider eigenmodes of a ring resonator made of a circular dielectric waveguide. Taking into account the polarization corrections, which are responsible for the interaction of polarization and orbital properties of electromagnetic waves (spin-orbit interaction of photons), results in fine splitting of the levels of scalar approximation. The basic features of this fine structure of the levels are similar to that of electron levels in an atom. Namely: 1) sublevels of the fine structure are defined by an additional quantum number: product of helicity of the wave and its orbital moment; 2) for a waveguide with a parabolic profile of the refractive index each level of the scalar approximation splits into $N$ sublevels ($N$ is the principal quantum number), while for any other profile it splits into 2 sublevels; 3) each level of the fine structure remains twice degenerated due to local axial symmetry of the waveguide. Numerical estimations show that the described fine splitting of levels may be observed in optic-fiber ring resonators.

## 1. INTRODUCTION

Spin-orbit interaction (SOI) phenomenon is well-known from the relativistic theory of particles with a spin [1]. Also known is the SOI of non-relativistic electrons in solid lattices (see [2] and references there). First of all this phenomenon leads to the fine splitting of energy levels of the particle; in particular it determines the fine structure of the electron levels in an atom [1]. Besides, SOI results in the transport effects as well. In this way a particle freely propagating in an external field experiences action of an additional forces connected with the SOI. Recently it was revealed that in some cases such forces have a topological nature and determine the so-called topological spin transport of particles (see, for instance, [3] and references there).

SOI is well investigated for massive particles. However, for a long time nothing was known about SOI of massless particles with a spin – for instance, photons. Meanwhile the photons are rather convenient objects for studying spin properties, since they are nothing else than the polarization effects for classical electromagnetic waves (see an example of a Berry phase that for the first time was predicted and observed for electromagnetic waves [4]).

The concept of the "spin-orbit interaction of photons" was for the first time proposed by Liberman and Zel'dovich in 1992 in the paper [5] (see also [6]). There the so-called optical Magnus effect was described (it was discovered by Zel'dovich et al. [7] shortly before this). The essence of the effect consists in the dependence of an electromagnetic wave (photon) trajectory on its polarization (a spin state or helicity) in locally isotropic inhomogeneous medium. This transport effect can be considered as a correction to classical Snell's law, according to which the refraction of an electromagnetic wave depends on its polarization [8] (see also [6] and references there). Papers [5–7] have pointed out the close connection of the optical Magnus effect, Berry phase, and SOI of photons. Recent papers [9–13] show that all these phenomena can be described by a single Berry gauge field in the momentum space and that the optical Magnus effect represents the topological spin transport of photons. At the same time, it was shown in papers [12,14,15] that SOI of electrons is a manifestation of the Berry gauge field in the momentum space as well. The latter fact determines the complete analogy between SOI of photons and electrons.

---
[*]E-mail: k_bliokh@mail.ru

Note that in all mentioned works SOI of photons manifests itself as a *transport* effect for freely propagating particles, but the possibility of *fine splitting of energy levels* was not mentioned anywhere (the reason, probably, is in the difficulty of the localization of a helical state of a photon). In the present work we for the first time give an example of the fine splitting of levels due to SOI of photons. For this we consider a ring dielectric resonator made of a circular multimode waveguide (exactly in such open waveguide the SOI of photons (optical Magnus effect) have been discovered for the first time in [7]). The main idea of the present paper is as follows.

Characteristics of the eigenmodes of a dielectric waveguide are practically polarization-independent, that is in the zero approximation the polarization degeneration takes place. Inhomogeneity of a refraction index profile in the waveguide leads to small polarization corrections, which are connected to the polarization and profile inhomogeneity [16]. The corrections describe SOI of photons and establish the dependence of the mode propagation constant on the polarization. Thus they remove the polarization degeneracy. In the case of an open circular waveguide these corrections result in azimuthal rotation of a speckle pattern depending on the polarization of radiation, this is transport optical Magnus effect [7]. If one closes the waveguide and make in this way a ring resonator with discrete values of the propagation constant, than obviously taking the polarization corrections into account will lead to fine splitting of levels. Bellow we will show that the main features of the fine structure of levels turn out to be quite similar to that of electron levels in an atom.

## 2. BASIC EQUATIONS

We will examine eigenmodes of a circular dielectric waveguide (fiber). Let us present the electric field of the modes as

$$\mathbf{E} = \mathbf{e}\exp(ik_\parallel z - i\omega t) + \text{c.c.} \; , \tag{1}$$

where $k_\parallel$ is the propagation constant of a given mode along the fiber, $\omega$ is the frequency, axis $z$ is directed along the waveguide axis, and $\mathbf{e}$ depends on the transverse coordinates only owing to the separation of variables in the medium with cylindrical symmetry. In paraxial approximation it is possible to consider vector $\mathbf{e}$ as practically transverse (orthogonal to $z$ axis). Then Maxwell equations for field (1) can be reduced to the form [16]

$$\left[\nabla_\perp^2 + k_\perp^2\right]\mathbf{e} = -\nabla_\perp\left(\mathbf{e}\nabla_\perp \ln n^2\right) . \tag{2}$$

Here $\nabla_\perp$ is the gradient in the plane of cross-section of the waveguide (i.e. transverse gradient), $n = n(\rho)$ is the refraction index ($\rho$ is the transverse radial coordinate in local cylindrical system of a fiber), and $k_\perp^2 = n^2 k_0^2 - k_\parallel^2$, where $k_0 = \omega/c$ is the wave number in vacuum. It can be seen that the term in the right-hand side of Eq. (2) is of the form of $\hat{\mathbf{A}}\nabla n$, where $\hat{\mathbf{A}}(\mathbf{p})$ is Berry gauge potential in the momentum space ($\mathbf{p} \propto \nabla$ is the momentum operator) [9–15]. As it is shown in [12,14,15] for electrons this term is nothing but SOI one.

In paraxial approximation the influence of the right-hand part of the equation (2) is small (even for waveguides with a step index) and it may be considered as perturbation [16]. Then in zero approximation Eq. (2) looks as a *scalar* wave equation. It means that field characteristics of the modes do not depend in any way on its polarization (a spin state), thus *polarizing degeneration* takes place. The correction in the right-hand side of Eq. (2) depends on wave polarization, and taking it into account removes the polarization degeneracy of the modes. In fact, the right-hand part of Eq. (2) describes SOI of an electromagnetic field (photons): it represents a combination of polarization and orbital characteristics of the field (compare with [6,8–13]).

Taking the right-hand side of Eq. (2) into account within the framework of the perturbation theory leads to dependence of the propagation constant of the mode on its polarization. It appears in the form of addition to $k_\parallel$ [16]:

$$\delta k_\parallel \approx -\frac{1}{k} \frac{\int (\nabla_\perp \mathbf{e})(\mathbf{e}^* \nabla_\perp \ln n) ds}{\int \mathbf{e}^* \mathbf{e}\, ds} \ . \qquad (3)$$

Here integrals are taken over the cross-section of the waveguide, the asterisk stands for complex conjugation and $k = n_0 k_0$ is the wave number at waveguide axis ($n_0 \equiv n(0)$).

As independent modes of a circular waveguide, the modes with right-hand and left-hand circular polarizations can be chosen. They correspond to the photon states with helicity $\sigma = \pm 1$ [1]:

$$\mathbf{e}_{l,m,\sigma} = \frac{1}{\sqrt{2}} \left( \mathbf{e}_x + i\sigma \mathbf{e}_y \right) \exp(im\varphi) F_{l,|m|}(\rho) \ . \qquad (4)$$

Here $\mathbf{e}_x$ and $\mathbf{e}_y$ are basis vectors of the corresponding Cartesian coordinate frame, while $\rho$ and $\varphi$ are coordinates in the cylindrical coordinate frame associated with the circular waveguide, $F_{l,|m|}(\rho)$ is the real radial function, $l = 1, 2, ...$ is the radial quantum number, and $m = 0, \pm 2, \pm 3 \pm 4, ...$ is the orbital quantum number (the orbital moment). Modes with $m = \pm 1$ cannot be presented in the form of Eq. (4) [6,16] and we do not consider them further.

## 3. LEVELS OF THE RING RESONATOR IN SCALAR APPROXIMATION

Let us consider a dielectric ring resonator, which constitutes a multimode circular waveguide (optical fiber, for instance) convoluted in a ring. The radius of the ring, $R$, is supposed to be much larger than the radius of the waveguide, $r$:

$$R \gg r \ , \qquad (5)$$

this allows one to not take into account the bend of the fiber at the local analysis of modes. Owing to this it is possible to think that eigenmodes of the resonator locally represent propagating modes of a straight circular waveguide with the additional condition of quantization being imposed.

For simplicity of analytical calculations we assume that the fiber has a parabolic profile of the square of refractive index:

$$n^2(\rho) = n_0^2 \left( 1 - 2\Delta \frac{\rho^2}{r^2} \right) . \qquad (6)$$

Here

$$\sqrt{\Delta} \ll 1 \qquad (7)$$

is the dimensionless parameter of height of the profile and condition (7) allows one to use paraxial approximation. Note also that a multimodeness condition of such fiber is

$$kr\sqrt{\Delta} \gg 1 \ . \qquad (8)$$

The characteristic equation for propagating modes in the fiber under consideration in the scalar approximation (i.e. without taking into account polarization corrections (3)) has the form [16]

$$\left( k^2 - k_\parallel^2 \right) r^2 = 2kr\sqrt{2\Delta}(2l + |m| - 1) > 0 \ . \qquad (9)$$

In an open waveguide $k$ has a continuous spectrum of values. In a ring resonator it is necessary to take into account the cyclic boundary conditions, which result in quantization of $k_\parallel$ and $k$. Indeed, by assuming that integer number of the wavelengths $2\pi/k_\parallel$ amount to the total length of the ring, $2\pi R$, we obtain

$$k_{\parallel q} = \frac{q}{R}, \qquad (10)$$

where $q = 1, 2, ...$ is the integer quantum number associated with the orbital quantization under motion along the ring. Note that multimodeness of the waveguide and paraxial approximation imply big values of $q$:

$$q \sim kR \gg 1. \qquad (11)$$

By substituting Eq. (10) in Eq. (9) we obtain the equation for $k$:

$$k^2 - 2k\frac{\sqrt{2\Delta}(2l+|m|-1)}{r} - \frac{q^2}{R^2} = 0. \qquad (12)$$

By solving Eq. (12) with respect to $k$, we have

$$k_{l,|m|,q} = \frac{\sqrt{2\Delta}(2l+|m|-1)}{r} + \sqrt{\frac{2\Delta(2l+|m|-1)^2}{r^2} + \frac{q^2}{R^2}}. \qquad (13)$$

This equation determines the structure of levels of the ring resonator without taking into account polarization corrections, i.e. spin-orbit interaction. Note that radial and azimuthal quantum numbers of the fiber are included into this expression only in combination $2l+|m|-1$. This allows one to introduce a single quantum number instead of them (analogue of the principal quantum number in atoms):

$$N = 2l+|m|-1 = 1, 2, ..., \quad k_{N,q} = \frac{\sqrt{2\Delta}N}{r} + \sqrt{\frac{2\Delta N^2}{r^2} + \frac{q^2}{R^2}}. \qquad (14)$$

It is worth noticing that the introduction of the principal quantum number $N$ is possible only in the case of the parabolic profile of the refractive index (6). Taking into account the two-fold polarization degeneracy, one can see that it determines a $2N$-fold degeneracy of each level. Indeed, exactly $N$ combinations of different $l$ и $m$ correspond to any given number $N$. In the general case of a non-parabolic refractive index's profile the levels are determined by quantum numbers $l, |m|, q$, and when $|m| > 1$ they are 4-fold degenerated (the polarization degeneracy and degeneracy of levels with opposite signs of $m$) [16].

Let us determine characteristic intervals between levels of the ring resonator as

$$\delta^{(N)}k = k_{N+1,q} - k_{N,q}, \quad \delta^{(q)}k = k_{N,q+1} - k_{N,q}. \qquad (15)$$

Conditions of the paraxial approximation and multimodeness, Eqs. (7) and (8), give in application to Eq. (14) the following condition (we assume $\Delta N \ll 1$):

$$\frac{NR\sqrt{\Delta}}{qr} \ll 1. \qquad (16)$$

In the first approximation with respect to small parameter (16) we approximately have from Eqs. (13)–(15)

$$k_{N,q} \approx \frac{q}{R} + \frac{\sqrt{2\Delta}N}{r}, \quad \delta^{(N)}k \approx \frac{\sqrt{2\Delta}}{r}, \quad \delta^{(q)}k \approx \frac{1}{R}. \qquad (17)$$

Hence the spectrum of the resonator is approximately equidistant. Depending upon the value of the characteristic parameter $R\sqrt{\Delta}/r$ both situations $\delta^{(N)}k \ll \delta^{(q)}k$, and $\delta^{(N)}k \gg \delta^{(q)}k$ can be realized (owing to Eq. (11) this parameter may be very different from the parameter (16)).

## 4. POLARIZATION SPLITTING AND FINE STRUCTURE OF LEVELS

In the case of a parabolic refractive index profile that we are considering, the radial functions of modes (4) equal [16]

$$F_{l,|m|}(\rho) = \tilde{\rho}^{|m|} L_{l-1}^{|m|}(V\tilde{\rho}^2) \exp(-V\tilde{\rho}^2/2) , \qquad (18)$$

where $\tilde{\rho} = \rho/r$, $L_{l-1}^{|m|}$ are generalized Laguerre polynomials, and $V = kr\sqrt{2\Delta}$ is the large waveguide parameter (see Eq. (8)). By substituting modes Eq. (4) with Eq. (18) in expression (3) it is possible to derive [16]

$$\delta k_{\| \sigma m} = -\frac{\Delta}{kr^2}(1+\sigma m) . \qquad (19)$$

Here a dependence on a new quantum number $\sigma m$, that is the product of the helicity and the orbital moment, has appeared. This number characterizes the spin-orbit interaction. Taking into account the polarization correction (19), we see that electric field of the modes, Eqs. (1), (4), takes the form

$$\mathbf{E}_{l,m,\sigma} = 2^{-1/2}(\mathbf{e}_x + i\sigma\mathbf{e}_y)\exp(im\varphi)F_{l,|m|}(\rho)\exp\left[i(k_\| + \delta k_{\|\sigma m})z - i\omega t\right] + \text{c.c.} . \qquad (20)$$

Thus, we have found the structure of modes that takes into account the polarization corrections for an open waveguide. Let us now turn to the ring resonator. Obviously the condition of quantization (resonance condition) Eq. (10) for a field (20) takes the form

$$k_{\|q} + \delta k_{\|\sigma m} = \frac{q}{R} . \qquad (21)$$

By substituting Eq. (21) into the characteristic equation (9) we obtain instead of Eq. (12)

$$k^2 - 2k\frac{\sqrt{2\Delta}N}{r} - \left(\frac{q}{R} - \delta k_{\|\sigma m}\right)^2 = 0 . \qquad (22)$$

It follows from here that taking into account the polarization correction (18) results in splitting of levels Eqs. (13), (14). Let us denote the shift of a level from its old value (14) as $\varepsilon$: $\tilde{k}_{N,q,\sigma m} = k_{N,q} + \varepsilon_{q,\sigma m}$. Then in the first approximation with respect to $\delta k_{\|l,|m|,\sigma m}$ we find from Eq. (22)

$$\varepsilon_{N,q,\sigma m} = -\frac{qr\delta k_{\|\sigma m}}{(k_{N,q}r - \sqrt{2\Delta}N)R} \approx -\frac{q\delta k_{\|\sigma m}}{k_{N,q}R} , \qquad (23)$$

where we have used Eqs. (7) and (8). Substitution of correction (19) and solutions (17) into Eq. (23) yields

$$\varepsilon_{q,\sigma m} \approx \frac{\Delta(1+\sigma m)R}{qr^2} . \qquad (24)$$

(To obtain expressions for the resonator's eigen frequencies and their shifts, obviously, one should just multiply corresponding values (13)–(15), (23), (24) by $c/n_0$.) For splitting to be considered as a fine one the following conditions have to be satisfied:

$$\varepsilon << \delta^{(N)}k, \delta^{(q)}k . \qquad (25)$$

First of them is satisfied automatically (see Eqs. (16) and (17)), whereas the second one requires the smallness of the parameter

$$\frac{R\sqrt{\Delta}}{r\sqrt{q}} << 1 . \qquad (26)$$

Expression (24) is the central result of this paper; this is the equation that determines the splitting and the fine structure of levels of the ring resonator under consideration. One can see from Eq. (24) that levels with $|m| > 1$ deviate in the opposite directions depending on the sign of $\sigma m$. (It is worth noticing that the empirical Hamiltonian of photon SOI in [5], as well as the Hamiltonian of SOI for a massive particle in a central-symmetric field [1], is also proportional to the scalar product of the spin and the orbital moment). Since levels of the resonator in zero approximation Eq. (14) are characterized by quantum numbers $N$ and $q$ only, than, as it was mentioned above, $N$ combinations with various $l$ and $m$ correspond to each scalar-approximation level, Eq. (14). Hence, it follows from Eq. (24) that *each level will be split into a multiplet of N sublevels*.

As was mentioned in Section 3, such situation takes place solely for parabolic profile of the waveguide's refractive index. In the general case every level of the scalar approximation with $|m| > 1$ is 4-fold degenerated and *spin-orbit interaction splits it into 2 sublevels*. The shift of the levels depends in this case on quantum numbers $l$ and $|m|$ too: $\varepsilon = \varepsilon_{l,|m|,q,\sigma m}$. Specific calculations for the levels splitting in the case of another, e.g. step-like, profile can be made by following the general scheme of the present paper and by using characteristic equations and polarization corrections listed in [16].

Note also that SOI correction (24) does not remove polarization degeneracy of levels completely: simultaneous change of the sign of the helicity and the orbital moment does not change a level. Thus each level remains *twice degenerate*. This double degeneracy is a consequence of local axial symmetry of the waveguide [17]. Indeed, instead of complex modes (4) $\mathbf{e}_{l,m,\sigma}$ and $\mathbf{e}_{l,-m,-\sigma} = \mathbf{e}^*_{l,m,\sigma}$ that correspond to the same levels of the resonator it is possible to use their superpositions (see [16]):

$$\mathbf{e}^{(1)}_{l,\sigma m} = \sqrt{2}\, \mathrm{Re}\, \mathbf{e}_{l,m,\sigma}\;, \quad \mathbf{e}^{(2)}_{l,\sigma m} = \sqrt{2}\, \mathrm{Im}\, \mathbf{e}_{l,m,\sigma}\;. \tag{27}$$

Modes $\mathbf{e}^{(1)}$ and $\mathbf{e}^{(2)}$ depend (excluding $l$) only on the product of the helicity and the orbital moment, $\sigma m$, and differ from each other (at $|m| > 1$) by a rotation with respect to the waveguide axis (see. Fig. 1).

Finally let us adduce numerical estimations for possible fine splitting of levels in the ring resonator made of multimode optical fiber. Let us take the characteristic parameters $\Delta \sim 10^{-2}$, $k \sim 10^7\,\mathrm{m}^{-1}$, $r \sim 10^{-4}\,\mathrm{m}$, $R \sim 1\,\mathrm{m}$. With these numbers we have $q \sim 10^7$, Eq. (11). By substituting these values in Eqs. (17) and (24), we obtain typical intervals between the levels of the resonator:

$$\delta^{(N)} k \sim 10^3\,\mathrm{m}^{-1}\;, \quad \delta^{(q)} k \sim 1\,\mathrm{m}^{-1}, \quad \delta^{(\sigma m)} k \sim 10^{-1}\,\mathrm{m}^{-1}\;, \tag{28}$$

were $\delta^{(\sigma m)} k = \Delta R / q r^2$ is the interval between two adjacent levels of the fine structure Eq. (24). Eqs. (28) shows that at the chosen values of parameters the characteristic interval of the fine structure is only one order less than the interval between adjacent zero-approximation levels of the resonator.

## 5. CONCLUSION

In conclusion, we have considered the influence of the spin-orbit interaction (SOI) on the structure of levels of a ring dielectric resonator. Examples of SOI of photons that were analyzed before produced additional *transport* effects only [5–13]. The possibility of splitting and formation of a fine structure of levels of the localized photon states is for the first time described in the present paper. It is quite similar to how it takes place for electrons in atoms. The main result is that taking into account the polarization corrections, which are responsible for SOI, results in partial removal of the degeneration and splitting of each level of the scalar approximation into a multiplet of equidistant sublevels ($N$ in the case of a parabolic profile of the waveguide index and 2 in other cases). Each sublevel is determined by an additional quantum number $\sigma m$, which is a product of the helicity and the orbital moment of the mode. Besides, each sublevel remains twice degenerated due to local axial symmetry of the waveguide. Thus *a close analogy with the spin-orbit interaction and the fine structure of electron levels in atoms* [1] *takes place*.

The numerical estimations given above for an optical fiber shows that the characteristic interval of the fine structure of levels makes $10^{-1} \div 10^{-4}$ characteristic intervals between levels in scalar approximation. This allows one to expect that the phenomenon can be observed experimentally. It may have an effect on the performance of resonant ring interferometers made of a multimode fiber [18]. In addition, splitting of the levels at the presence of rotation or additional Faraday elements in a circuit may cause interesting structures similar to driving quantum multilevel systems [19].

**ACKNOWLEDGMENTS**

The work was partially supported by INTAS (Grant No. 03-55-1921) and by Ukrainian President's Grant for Young Scientists GP/F8/51.

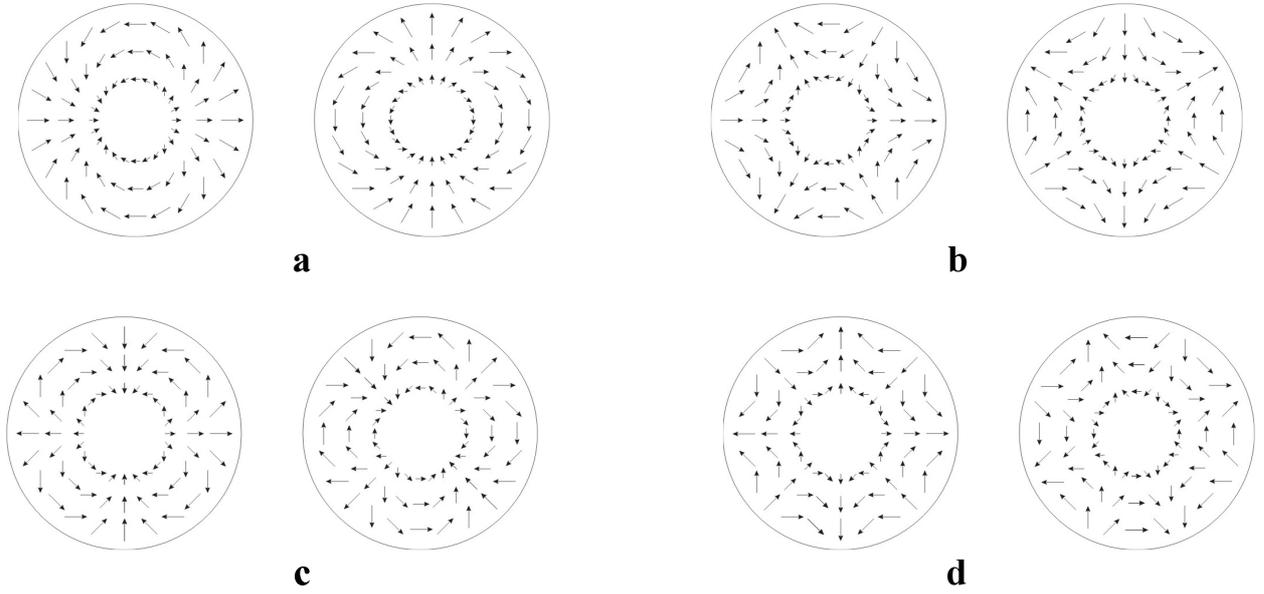

**Fig. 1.** Distribution of the directions of electric field in the cross-section of a circular waveguide for modes $\mathbf{e}^{(1)}$ (left-hand picture in each pair) and $\mathbf{e}^{(2)}$ (right-hand pictures), Eq. (27). Pairs (a), (b), (c) and (d) correspond to values $\sigma m$ 2, $-2$, 3 and $-3$ respectively. Dependence of the field intensity on the radial coordinate $\rho$ is not considered here. One can see that when making a complete round of the waveguide cross-section along azimuth coordinate $\varphi$ the electric field vector makes precisely $\sigma m$ revolutions (positive revolutions are clockwise and negative are anticlockwise).